# WiFi Motion Detection: A Study into Efficacy and Classification

Sadhana Lolla, Amy Zhao
Johns Hopkins University Applied Physics Laboratory
sadhana.lolla@jhuapl.edu, amy.zhao@jhuapl.edu

*Abstract* - **WiFi and security pose both an issue and act as a growing presence in everyday life. Today's motion detection implementations are severely lacking in the areas of secrecy, scope, and cost. To combat this problem, we aim to develop a motion detection system that utilizes WiFi Channel State Information (CSI), which describes how a wireless signal propagates from the transmitter to the receiver. The goal of this study is to develop a real-time motion detection and classification system that is discreet, cost-effective, and easily implementable. The system would only require an Ubuntu laptop with an Intel Ultimate N WiFi Link 5300 and a standard router. The system will be developed in two parts: (1) a robust system to track CSI variations in real-time, and (2) an algorithm to classify the motion. The system used to track CSI variance in real-time was completed in August 2018. Initial results show that introduction of motion to a previously motionless area is detected with high confidence. We present the development of (1) anomaly detection, utilizing the moving average filter implemented in the initial program and/or unsupervised machine learning, and (2) supervised machine learning algorithms to classify a set of simple motions using a proposed feature extraction methods. Lastly, classification methods such as Decision Tree, Naive Bayes, and Long Short-Term Memory can be used to classify basic actions regardless of speed, location, or orientation.**

## INTRODUCTION

In the modern world, WiFi is becoming more ubiquitous by the day. The presence of WiFi has such a profound impact on people's daily lives that there have been discussions of internet access becoming a human right. Such a powerful tool should be harnessed and utilized. We propose creating a device that can monitor the WiFi state in a given location so that motion can be detected, classified, and monitored.

There are countless implementations for such a motion sensor. Applications include: (1) clandestine operations, where the primary objective is security and protecting the integrity of one's base of operations, (2) healthcare, where the elderly and ill patients may be monitored for safety and peace of mind, (3) educational institutions, where students may be kept safe through the detection of unauthorized intruders, and (4) professional settings, where businesses can analyze traffic and generate methods of attracting customers.

Numerous problems arise when the aforementioned applications are implemented with current state-of-the-art motion detection systems. Many sensors utilize substantial equipment, making them easily detectable and leaving a large optical footprint. This poses a problem for operations where secrecy is the key for success. Another problem that arises is the need for line-of-sight (LOS). In modern solutions, multiple motion sensors are needed in order to encompass the full scope of monitoring. Blind spots in the LOS allows for trespassers to easily avoid detection, jeopardizing people and information [1]. In professional or commercial environments, people may feel that the excessive equipment is invasive in their daily lives. Additionally, such equipment is often expensive and difficult to maintain, compounding the costs of security in the modern world.

WiFi motion detection is easily able to resolve these problems. A WiFi based motion detector does not rely on lighting, radar, or multiple devices. Large and complex areas can be monitored with a single router that sends data to a receiving laptop as WiFi propagates through the surrounding area. LOS is no longer a concern as WiFi waves circulate throughout the entire space, not just the specific area covered by traditional LOS sensors. WiFi routers are common everyday objects that: (1) have no optical footprint, so secrecy can be achieved, and (2) removes the need to setup new equipment since they usually exist as part of the building's infrastructure.

## LITERATURE REVIEW

### I. Applications of WiFi CSI

Many previous studies have examined the use of WiFi CSI as it relates to human activity recognition. CSI has previously been used to localize motion [2], detect the presence of a person in a room/building [3], and count the number of people in a crowd [4]. Other applications include using CSI to detect keystrokes and using specialized directional antennas to obtain the CSI variations caused by lip movements during speaking [5]. While received signal strength (RSS) has been used for the same purposes, CSI is selected because RSS cannot capture the real changes in the signal due to the movement of the person. By obtaining the CSI for each subcarrier in orthogonal frequency division multiplexing (OFDM) systems, the observed channel dynamics will display diversity, as opposed to RSS, which cannot capture the change at certain frequencies [6].

*II. Previous Implementations*

The open-source tool developed by Daniel Halperi at the University of Washington was utilized to collect CSI data from the WiFi router. This tool does not provide a real-time implementation for motion detection applications, and the data must be visualized separately. Furthermore, a graphical user interface (GUI) is not built into the application [7]. Other real-time applications include one by Chao Cai [8], and one by Bingxian Lu [9]. However, the application by Cai et al. has significant packet loss through the linear processing of the packets and ineffective data storage [8]. In this paper, we utilize the modified WiFi driver by Halperi et al. as a backbone for our implementation.

## BACKGROUND

CSI (Channel State Information) refers to the channel properties of any wireless communication link. It describes how a signal propagates from the transmitter to the receiver. CSI can be categorized into instantaneous CSI (short-term) and statistical CSI (long-term). In this implementation, WiFi CSI reveals a set of channel measurements that describe amplitudes and phases of each of 30 subcarriers as given by the equation (1):

$$H(f_k) = \|H(f_k)\| e^{j \sin(\angle H)} \qquad (1)$$

Where $H(f_k)$ is the CSI value at the subcarrier with central frequency of $f_k$, and $\angle H$ is the phase.

WiFi NICs continuously monitor variations in the wireless channel using CSI. Let $X(f,t)$ and $Y(f,t)$ be the frequency domain representations of transmitted and received signals, respectively, with carrier frequency f. The two signals are related by the expression: $Y(f,t) = H(f,t) \times X(f,t)$, where $H(f,t)$ is the complex valued channel frequency response (CFR) for carrier frequency $f$ measured at time $t$. CSI measurements basically contain these CFR values. Here, $NRx$ equals the total number of antennas used to receive the packet by this NIC and $NTx$ represents the number of space/time streams transmitted. As CSI is measured on 30 selected OFDM subcarriers for a received 802.11 frame, each CSI measurement contains 30 matrices with dimensions $NTx \times NRx$. [1]

## METHODOLOGY

The development was split into two segments: (1) developing a better real-time processing system for the CSI tool developed by Halperi et. al, and (2) classifying the motion.

Creating a practical motion detection system entails four main tasks in order to plot CSI variance in real time. This was accomplished through Python's multiprocessing module which allowed the simultaneous execution of four processes. These processes addressed the following problems: (1) how to ingest incoming packets received from the router, (2) how to parse the packets into meaningful data, (3) how to store data for future recall purposes, and (4) how to plot CSI amplitude and variance in a time series.

*I. Collecting EM Waves via an Intel WiFi Chip:*

A TCP server/client was implemented to ingest incoming packets. In previous implementations, packets were dropped because the packet had to be a certain size, as there were only two antennae in use. In our implementation, we received packets from all three antennae of the router, so we could process packets of any size. Packet loss occurred in previous implementations because data was not processed before the next packet arrived. We utilize a queue because the first-in-first-out (FIFO) processing scheme prevents premature processing from occurring.

*II. Processing Data*

Once the data were popped off the queue, the algorithm split the data into bytes as shown in Figure I.

| Header | | | | | | | | |
|---|---|---|---|---|---|---|---|---|
| Field | Time stamp | bfee count | Nrx | Ntx | rssi_a | rssi_b | rssi_c | noise | agc |
| Byte Offset | 0-3 | 4-6 | 7 | 8 | 9 | 10 | 11 | 12 | 13 |
| Type | int32 | u16 | u16 | u16 | u16 | u16 | u16 | int8 | u16 |

| Header | | | Payload |
|---|---|---|---|
| Field | antenna_sel | length | rate | payload |
| Byte Offset | 14 | 16, 17 | 18, 19 | 20-n |
| Type | u16 | u16 | u16 | u8 |

FIGURE I
PACKET OUTPUT DATA

Then the data were converted into a matrix of size NTx × NRx × 30 because there are 30 channels. Here, NRx equals the total number of antennas used to receive the packet by this NIC and NTx represents the number of space/time streams transmitted. Since the amplitude and variance can be analyzed using just one CSI channel, the data from the first subcarrier was parsed. The NTx × NRx × 1 matrix was then converted into the packet structure as shown above. The CSI data is stored in the payload as a complex number. The Euclidean norm of this complex number was then passed to processes 3 and 4, along with a scaling factor to obtain data in absolute units. This scaling factor was computed by combining other data in the packet such as RSSI (relative received signal strength) and AGC (automatic gain control) [7].

*III. Database Storage*

Considering the volume of data received from the router, a MySQL database was initialized to store the parsed packets. Since a goal of this project was discretion, MySQL offered the advantage of data security. When the database was initialized, login credentials were set to include a layer of insurance. Additionally, MySQL offered unrivaled scalability, which helped maintain the database, given the amount of data being written into the database. The database includes four columns: packet ID, timestamp, CSI amplitude, and CSI variance. Implementation of the MySQL database into the fundamental Python program was done through the MySQL Connector/Python. This allowed for seamless communication between the multiprocessing thread and the database.

*IV. Plotting Time Series Data in Real-Time*

The final process of the real-time implementation is the display in real-time. The program plotted through matplotlib using pyplot, which contains functions meant to emulate those in MATLAB. Given that the original range of the data is unknown, axes were automatically formatted as data points were graphed. A five-point moving average filter was later implemented to smoothen the incoming data, filtering out noisy components that were not statistically significant from the data.

### TESTING METHODOLOGY

Initial testing in the anechoic chamber showed that the real-time system collected data similar to the system implemented by Halperi et al. The real-time program was left running in the anechoic chamber at APL for two minutes, after which a person opened the door to the chamber and walked back and forth three times. The same was done with the system implemented by Halperi et al and the data from this application was visualized at a later date. Utilizing the anechoic chamber allowed us to control the RF traffic in our testing location, ensuring the data was not corrupted by some other source.

The same methodology was used to conduct trials outside of the anechoic chamber to compare the differences in amplitude and variance.

In another test, the real-time system was left running inside the anechoic chamber with no motion, and the only motion would be to start or stop the trial. The same was done with the system developed by Halperi et al. This process was conducted inside and outside the anechoic chamber. Results from one trial are shown in figures 2, 3, 4, and 5.

### INITIAL RESULTS

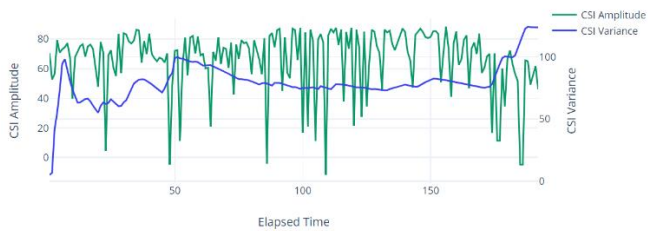

FIGURE 2
REAL-TIME CSI AMPLITUDE AND VARIANCE OUTSIDE ANECHOIC CHAMBER

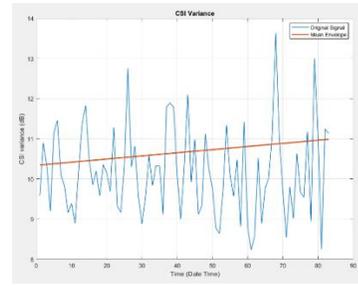

FIGURE 3
HALPERI ET AL IMPLEMENTATION CSI AMPLITUDE AND VARIANCE OUTSIDE ANECHOIC CHAMBER

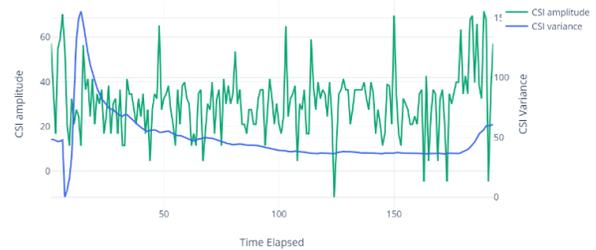

FIGURE 4
REAL-TIME CSI AMPLITUDE AND VARIANCE INSIDE ANECHOIC CHAMBER

The data in figures 2 and 3 were collected outside of the anechoic chamber, which accounts for the larger variations in amplitude than the data collected within the chamber (Figures 4 and 5). Because both implementations were not running simultaneously, the graphs do not show the exact same data. It should be noted that in Figure 4, the program was collecting data at 100Hz, whereas in the other trials, the program was collecting data at 1Hz.

In figures 1 and 2, the router and laptop were placed in close proximity with an absence of motion in the room, which is demonstrated by the relatively constant variance. In figures 3 and 4, the data were collected in the anechoic chamber with the same setup, so it was hypothesized that there would be constant variance except for the intervals in which we were in the chamber to start or stop the algorithm. This is true for both figures 3 and 4, where the variance changes at the beginning and at the end of the time interval, as there was human activity in the anechoic chamber.

### CONCLUSION

Initial results are promising because the data and graphs generated from both programs are similar, indicating that this may be a viable real-time application for motion detection. The data from the new real-time implementation shows that the implementation can clearly differentiate between the presence and absence of human activity, exhibiting great promise as a cost-effective, accurate motion detection system.

### FUTURE WORK

*I. Anomaly Detection Using K-Means Reconstruction*

*K-Means:* Anomaly detection through k-means uses data that is first split into overlapping segments in order to obtain instances of each data shape with a variety of horizontal translations. Applying a window function to the data forces each segment to start and end at zero which will ultimately allow for a continuous reconstruction. The k-means algorithm is provided by Python's scikit-learn library. Once the data segments are clustered, the data is reconstructed using the library of learned shapes. The cluster centroid that best matches the segment will be used for reconstruction and the reconstruction segments will be joined together. When anomalies are introduced, it produces a shape that is unable to be reconstructed from the learned shape library, producing an easily discernible error.

*II. Motion Classification using Supervised Machine Learning Techniques*

*Dataset:* A large dataset of CSI movements is located at [11], where the authors of the study sought to classify six different types of movements: laying down, sitting, standing up, running, walking, and picking up an object. This dataset can be used to train the machine learning algorithms described below.

*Feature Selection:* Feature extraction in previous studies has been accomplished using DWT (Discrete Wavelet Transform), which can extract features as a function of time.

*Classification using Decision Tree:* Decision Tree is a simple supervised machine learning algorithm that can be implemented using the sci-kit learn module in Python. Decision trees are known as "white-box" algorithms because every decision can be explained with a series of if-else statements [12]. For each attribute in a dataset, a decision tree creates a node and the most important attribute is placed at the root node. When predicting an outcome, the algorithm traverses the path taken by the decision tree until a leaf node is reached. Decision tree algorithms are also relatively fast and efficient.

*Classification with Naive Bayes:* The Naive Bayes Classifier relies on Bayes Theorem, which is based on conditional probability. We assume that the features are independent. The benefit of using this technique is that is requires only a small amount of training data. Given the training data, the Naive Bayes algorithm compares a group of data points and calculates the probability that they are a certain feature. The feature with the highest calculated probability will be used to classify the data points.

*Classification with Long Short-Term Memory (LSTM):* The LSTM algorithm is a recurrent neural network that is able to pre-process data and hold temporal state information of the activity, giving it the potential to distinguish similar activities [6]. LSTMs are designed with multiple neural network layers. The model learns to extract features from sequences of data and how to map these features to different activity types. Defining the model requires a three-dimensional input with samples, time steps, and features.

The output for the model will be a six-element vector containing the probability of a given window pertaining to each of the types of motion. The model will have a single LSTM hidden layer, a dropout layer intended to reduce overfitting, a dense, fully connected layer to interpret the features extracted from the hidden layer, and a final output layer to make predictions. Once the model is fit, the test data set will be used to determine its accuracy.

ACKNOWLEDGEMENT

Sincere thanks to Sean Wang, Emily Brown, J.R. Parsons, and Colleen D'Agrosa at JHU APL for their support.

AUTHOR INFORMATION

**Sadhana Lolla,** Research Intern, Poolesville High School/Johns Hopkins Applied Physics Laboratory.

**Amy Zhao,** Research Intern, River Hill High School/Johns Hopkins Applied Physics Laboratory.